\documentclass[aps,showpacs,twocolumn,amssymb]{revtex4}
\usepackage{amssymb}
\usepackage{graphicx}
\usepackage{color}
\usepackage{textcomp}
\usepackage{ulem}
\begin{document}

\title{Reduction of (pseudo-)Critical Temperatures of Chiral Restoration and Deconfinement Phase Transitions in a Magnetized PNJL Model}
\author{Shijun Mao}
\affiliation{School of Science, Xi'an Jiaotong University, Xi'an, Shaanxi 710049, China}

\begin{abstract}
We investigate the chiral restoration and deconfinement phase transitions under external magnetic field in frame of a Pauli-Villars regularized PNJL model. A running Polyakov loop scale parameter $T_0(eB)$ is introduced to mimic the reaction of the gluon sector to the presence of magnetic fields. It is found that a decreasing $T_0(eB)$ with magnetic fields can realize the inverse magnetic catalysis phenomena of chiral condensates of $u$ and $d$ quarks, increase of Polyakov loop and the reduction of (pseudo-)critical temperatures of chiral restoration and deconfinement phase transitions.
\end{abstract}

\date{\today}
\maketitle

\section{Introduction}
The study on Quantum Chromodynamics (QCD) phase structure is recently extended to including external electromagnetic fields, motivated by the strong magnetic field in the core of compact stars and in the initial stage of relativistic heavy ion collisions~\cite{review0,review1,review2,review3,review4,review5,lattice1,lattice2,lattice4,lattice5,lattice6,lattice7,lattice9,lattice8,fukushima,mao,kamikado,bf1,bf13,bf2,bf3,bf5,bf51,bf52,bf8,bf9,bf11,db1,db2,db3,db5,db6,pnjl1,pnjl2,pnjl3,pnjl4,pqm,ferr1,ferr2,mhuang,meimao1,t0effect}. Chiral restoration and deconfinement are the two most important QCD phase transitions at finite temperature and magnetic field.

From recent lattice QCD simulations with a physical pion mass~\cite{lattice1,lattice2,lattice4,lattice5,lattice6,lattice7,lattice9}, the chiral condensates of light quarks ($u$ and $d$) are enhanced by magnetic fields in vacuum, which is the magnetic catalysis phenomena, but they are suppressed by magnetic fields in high temperature region, which is the inverse magnetic catalysis phenomena. However, the chiral condensates of $s$ quark and $u,\ d$ quarks with heavy current masses show magnetic catalysis phenomena in the whole temperature region~\cite{lattice7,lattice8}. The (pseudo-)critical temperatures of the chiral restoration phase transition of $u,\ d,\ s$ quarks drop down with increasing magnetic field. Meanwhile, lattice simulations report that the renormalized Polyakov loop increases with magnetic fields and the transition temperature of deconfinement decreases as the magnetic field grows~\cite{lattice1,lattice2,lattice4,lattice5,lattice9}.

On analytical side, in the presence of a uniform external magnetic field ${\bold B} =B{\it {{\bold e}_z}}$, the energy dispersion of quarks takes the form $E=\sqrt{p_z^2+2|QB|l+m^2}$ with the momentum $p_z$ along the direction of magnetic field and the Landau level $l=0,1,2,...$~\cite{landau}. Due to this fermion dimension reduction, almost all model calculations at mean field level present the magnetic catalysis effect of chiral condensates in the whole temperature region and the increasing (pseudo-)critical temperatures for chiral restoration and deconfinement phase transitions under external magnetic field, see review~\cite{review0,review1,review2,review4,review5} and the references therein. How to explain the inverse magnetic catalysis phenomena and the reduction of (pseudo-)critical temperatures for chiral restoration and deconfinement phase transitions is open questions. Many scenarios are proposed~\cite{fukushima,mao,kamikado,bf1,lattice9,bf13,bf2,bf3,bf5,bf51,bf52,bf8,bf9,bf11,db1,db2,db3,db5,db6,pnjl1,pnjl2,pnjl3,pnjl4,pqm,ferr1,ferr2,mhuang,meimao1,t0effect}, such as magnetic inhibition of mesons, sphalerons, gluon screening effect, weakening of strong coupling, and anomalous magnetic moment.

In this paper, we will revisit the magnetic field effect on chiral restoration and deconfinement phase transitions in terms of two-flavor and three-flavor Nambu-Jona-Lasinio model with Polyakov loop (PNJL). According to the lattice QCD analysis~\cite{lattice9}, the interaction between Polyakov loop and sea quarks may be important for the mechanism of inverse magnetic catalysis and the reduction of transition temperatures. We consider a magnetic field dependent Polyakov loop scale parameter $T_0(eB)$ to mimic the reaction of the gluon sector to the presence of magnetic fields. This type of procedure on parameter $T_0$ had already been proposed in different context, such as in two-flavor Polyakov loop extended quark-meson model (PQM)~\cite{t0effect}, and three-flavor (entangled) PNJL model~\cite{pnjl3}. PQM model~\cite{t0effect} reported that the reduction of (pseudo-)critical temperature of chiral restoration happens only in weak magnetic field region. The PNJL model with magnetic field independent regularization scheme cannot reproduce the decreasing (pseudo-)critical temperatures for chiral restoration and deconfinement, but entangled PNJL model can realize it~\cite{pnjl3}. With a Pauli-Villars regularization scheme in our two-flavor and three-flavor PNJL model, we find that it is possible to account for the inverse magnetic catalysis phenomena of chiral condensates of $u,\ d$ quarks, increase of Polyakov loop and the reduction of (pseudo-)critical temperatures of chiral restoration and deconfinement phase transitions, when introducing a fast decreasing $T_0(eB)$.

The paper is organized as follows. Section \ref{2fframe} devotes to the two-flavor PNJL model, which introduces the framework in Sec.\ref{2fframe}(A) and discusses the numerical results in Sec.\ref{2fframe}(B). Similar study is extended to three-flavor PNJL model in Sec.\ref{3fframe}. Finally, we give the summary in Sec.\ref{summary}.

\section{two-flavor PNJL model}
\label{2fframe}

\subsection{Theoretical Framework}

The two-flavor PNJL model in external electromagnetic fields is defined through the Lagrangian density~\cite{pnjl5,pnjl6,pnjl7,pnjl8,pnjl9,pnjl10,pnjl12,t044,t041},
\begin{equation}
\label{pnjl}
{\cal L} = \bar\psi(i\gamma_\mu D^\mu-\hat{m}_0)\psi + {G\over 2}\left[\left(\bar\psi\psi\right)^2 + \left(\bar\psi i \gamma_5 {\bf \tau}\psi\right)^2 \right]-{\cal U}(\Phi,\bar\Phi).
\end{equation}
The covariant derivative $D^\mu=\partial^\mu+i Q A^\mu-i {\cal A}^\mu$ couples quarks to the two external fields, the magnetic field ${\bf B}=\nabla\times{\bf A}$ and the temporal gluon field  ${\cal A}^\mu=\delta^\mu_0 {\cal A}^0$ with ${\cal A}^0=g{\cal A}^0_a \lambda_a/2=-i{\cal A}_4$ in Euclidean space. The gauge coupling $g$ is combined with the SU(3) gauge field ${\cal A}^0_a(x)$ to define ${\cal A}^\mu(x)$, and $\lambda_a$ are the Gell-Mann matrices in color space. In this work, we consider magnetic field ${\bf B}=(0, 0, B)$ along the $z$-axis by setting $A_\mu=(0,0,x B,0)$ in Landau gauge, which couples with quarks of electric charge $Q=diag(Q_u, Q_d)=diag(2e/3,-e/3)$. $\hat{m}_0=diag(m^u_0, m^d_0)=diag(m_0,m_0)$ is the current quark mass matrix in flavor space, which controls the explicit breaking of chiral symmetry. For the chiral section in the Lagrangian, $G$ is the coupling constant in the scalar and pseudo-scalar channels, which determines the spontaneous breaking of chiral symmetry. The Polyakov potential describing deconfinement at finite temperature reads
\begin{equation}
\label{polyakov}
{{\cal U}(\Phi,{\bar \Phi})\over T^4} = -{b_2(t)\over 2} \bar\Phi\Phi -{b_3\over 6}\left({\bar\Phi}^3+\Phi^3\right)+{b_4\over 4}\left(\bar\Phi\Phi\right)^2,
\end{equation}
where $\Phi$ is the trace of the Polyakov loop $\Phi=\left({\text {Tr}}_c L \right)/N_c$, with $L({\bf x})={\cal P} \text {exp}[i \int^\beta_0 d \tau A_4({\bf x},\tau)]= \text {exp}[i \beta A_4 ]$ and $\beta=1/T$, the coefficient $b_2(t)=a_0+a_1 t+a_2 t^2+a_3 t^3$ with $t=T_0/T$ is temperature dependent, and the other coefficients $b_3$ and $b_4$ are constants.

The order parameter to describe chiral restoration phase transition is the chiral condensate $\sigma=\langle\bar\psi\psi\rangle$ or the dynamical quark mass $m=m_0-G \sigma $~\cite{njl1,njl2,njl3,njl4,njl5}. $\Phi$ is considered as the order parameter to describe the deconfinement phase transition, which satisfies $\Phi \rightarrow 0$ in confined phase at low temperature and $\Phi \rightarrow 1$ in deconfined phase at high temperature~\cite{pnjl5,pnjl6,pnjl7,pnjl8,pnjl9,pnjl10,pnjl12,t044,t041}. In mean field approximation, the thermodynamic potential at finite temperature and magnetic field contains the mean field part and quark part
\begin{eqnarray}
\label{omega1}
\Omega_{\text {mf}} &=&{\cal U}(\Phi,{\bar \Phi})+ \frac{(m-m_0)^2}{2 G}+\Omega_q,\\
\Omega_q &=& - \sum_{f,n}\alpha_n \int \frac{d p_z}{2\pi} \frac{|Q_f B|}{2\pi} \big[3E_f\nonumber\\
&+& T\ln\left(1+3\Phi e^{-\beta E_f}+3{\bar \Phi}e^{-2\beta E_f}+e^{-3\beta E_f}\right)\nonumber\\
&+& T\ln\left(1+3{\bar \Phi} e^{-\beta E_f}+3{ \Phi}e^{-2\beta E_f}+e^{-3\beta E_f}\right)\big],\nonumber
\end{eqnarray}
with spin factor $\alpha_n=2-\delta_{n0}$ and quark energy $E_f=\sqrt{p^2_z+2 n |Q_f B|+m^2}$ of flavor $f=u,d$ and Landau level $n$.

The ground state is determined by minimizing the thermodynamic potential,
\begin{eqnarray}
\label{gapeqs}
\frac{\partial\Omega_{\text {mf}}}{\partial m}=0,\ \frac{\partial\Omega_{\text {mf}}}{\partial \Phi}=0,\ \frac{\partial\Omega_{\text {mf}}}{\partial {\bar \Phi}}=0,
\end{eqnarray}
which lead to three coupled gap equations for the order parameters $m$, $\Phi$ and ${\bar \Phi}$. Note that there is $\Phi=\bar\Phi$ at vanishing baryon density. Therefore, in our current work, we only need to solve two coupled gap equations,
\begin{eqnarray}
\label{gapeqst}
\frac{\partial\Omega_{\text {mf}}}{\partial m}=0,\ \frac{\partial\Omega_{\text {mf}}}{\partial \Phi}=0.
\end{eqnarray}

Because of the contact interaction among quarks, NJL models are non-renormalizable, and it is necessary to introduce a regularization scheme to remove the ultraviolet divergence in momentum integrations. In this work, we take a Pauli-Villars regularization~\cite{mao}, which is gauge invariant and can guarantee the law of causality at finite magnetic field. The three parameters in the two-flavor NJL model, namely the current quark mass $m_0=5$ MeV, the coupling constant $G=7.79$ GeV$^{-2}$, and the mass parameter $\Lambda=1127$ MeV are fixed by fitting the chiral condensate $\langle \bar \psi \psi \rangle =(-250\ \text{MeV})^3$, pion mass $m_{\pi}=134$ MeV, and pion decay constant $f_\pi=93$ MeV in vacuum. For the Polyakov potential, its temperature dependence is fitted from the lattice simulation, and the parameters are chosen as~\cite{pnjl6} $a_0=6.75$, $a_1=-1.95$, $a_2=2.625$, $a_3=-7.44$, $b_3=0.75$, $b_4=7.5$ and $T_0=270$ MeV.


\subsection{Numerical Results}

\begin{figure}[hb]
\centering
\includegraphics[width=7.5cm]{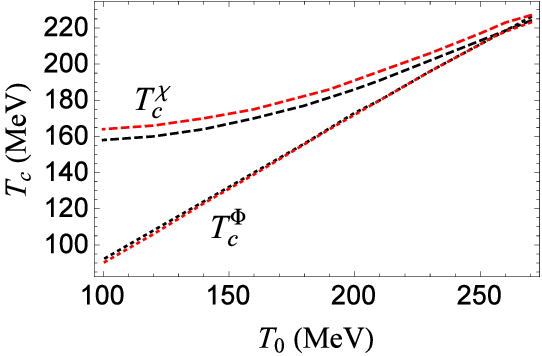}
\caption{The (pseudo-)critical temperatures $T_c^\chi, T_c^\Phi$ of chiral restoration and deconfinement phase transitions as functions of parameter $T_0$ with fixed magnetic field $eB/m^2_\pi=0$ (black lines) and $eB/m^2_\pi=10$ (red lines).}
\label{figtct0}
\end{figure}
\begin{figure}[hb]
\centering
\includegraphics[width=7.5cm]{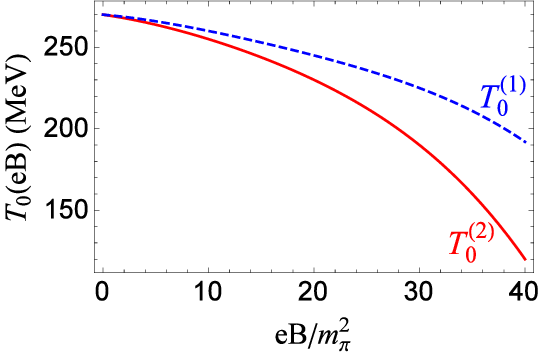}\\
\includegraphics[width=7.5cm]{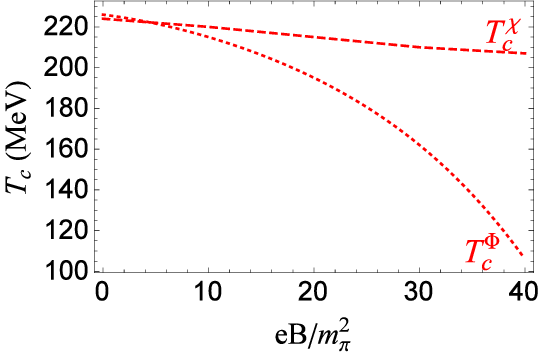}
\caption{(Upper panel) Two examples of magnetic field dependent parameter $T_0$ in Polyakov potential, $T^{(1)}_0(eB)$ (blue line) and $T^{(2)}_0(eB)$ (red line). (Lower panel) the (pseudo-)critical temperatures $T_c^\chi, T_c^\Phi$ for chiral restoration and deconfinement phase transitions as functions of magnetic fields with $T^{(2)}_0(eB)$.}
\label{figtct0eb270}
\end{figure}
\begin{figure}[hb]
\centering
\includegraphics[width=7.5cm]{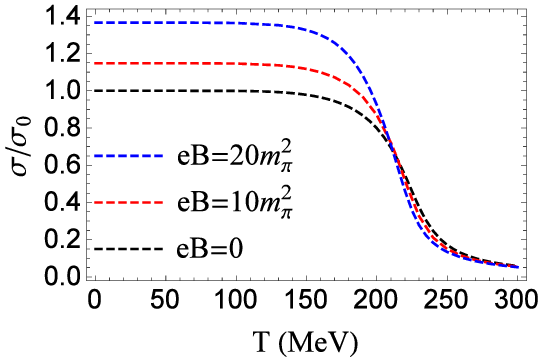}\\
\includegraphics[width=7.5cm]{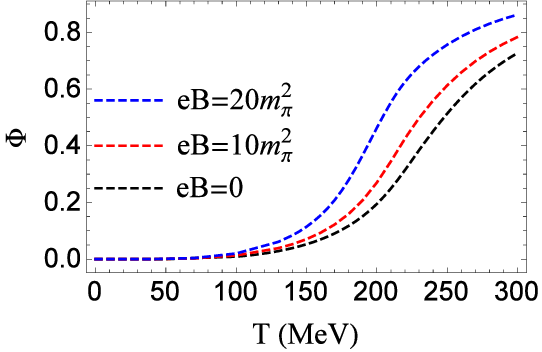}
\caption{The order parameters, chiral condensate $\sigma/\sigma_0$ (upper panel) and Polyakov loop $\Phi$ (lower panel) as functions of temperature with fixed magnetic field $eB/m^2_\pi=0,\ 10,\ 20$ and $T^{(2)}_0(eB)$. Here, $\sigma_0$ is the chiral condensate in vacuum with vanishing temperature, density and magnetic field.}
\label{figmqphiimc}
\end{figure}

As we know, the model parameter $T_0=270$ MeV is the critical temperature for the deconfinement phase transition in the pure gauge Polyakov loop model, and the inclusion of dynamical quarks will lead to a decrease of $T_0$~\cite{t044,t041}. External magnetic field makes the change on the quark properties, so that it will also alter the value of parameter $T_0$. This modification in the Polyakov potential will affect the chiral restoration and deconfinement phase transitions under external magnetic fields. In this work, we introduce a magnetic field dependent parameter $T_0(eB)$ to mimic the reaction of gluon sector to the presence of magnetic fields, and consider its effect on the chiral restoration and deconfinement phase transitions.

In Fig.\ref{figtct0}, we plot the (pseudo-)critical temperatures $T_c^\chi, T_c^\Phi$ for chiral restoration and deconfinement phase transitions as functions of parameter $T_0$ with fixed magnetic field $eB/m^2_\pi=0$ (black lines) and $eB/m^2_\pi=10$ (red lines). At finite temperature and/or magnetic field, the chiral restoration and deconfinement phase transitions are smooth crossover. The (pseudo-)critical temperature $T_c^\chi$ is defined by the condition $\frac{\partial^2 m}{\partial T^2}=0$ and $T_c^\Phi$ is defined by $\frac{\partial^2 \Phi}{\partial T^2}=0$. On one side, with fixed magnetic fields, the (pseudo-)critical temperatures $T_c^\chi$ and $T_c^\Phi$ decrease as $T_0$ decreases. Moreover, $T_c^\Phi$ decreases faster than $T_c^\chi$, which results in a larger separation between them at smaller value of $T_0$. The reason is the direct influence of Polyakov loop parameter $T_0$ on the gluon sector and indirect influence on the quark sector. On the other side, when fixing $T_0$, $T_c^\chi$ is higher with stronger magnetic fields, which is the typical result of mean field calculations in effective models~\cite{review0,review1,review2,review4,review5}. $T_c^\Phi$ is not sensitive to the magnetic field, because there is no direct interaction between the magnetic field and gluon field. Considering the effects of parameter $T_0$ and external magnetic field $eB$ on chiral restoration and deconfinement phase transitions, we can expect that with a fast decreasing $T_0$ under magnetic fields, it is possible to observe the reduction of (pseudo-)critical temperatures $T_c^\chi$ and $T_c^\Phi$ as the magnetic field grows.

Figure \ref{figtct0eb270} shows two explicit examples of magnetic field dependent $T_0(eB)$. In case of $T^{(1)}_0(eB)$, see the blue line in the upper panel, we obtain a constant (pseudo-)critical temperature of chiral restoration $T_c^\chi$ with different magnetic fields. It should be mentioned that due to separation between $T_c^\chi$ and $T_c^\Phi$, it is not possible to obtain constant $T_c^\chi$ and $T_c^\Phi$ simultaneously. When we have a faster decreasing $T^{(2)}_0(eB)$, see the red line in the upper panel, both (pseudo-)critical temperatures $T_c^\chi,\ T_c^\Phi$ of chiral restoration and deconfinement phase transitions (see the lower panel) decrease with increasing magnetic fields, which show similar trend as LQCD results~\cite{lattice1,lattice2,lattice4,lattice5,lattice6,lattice7,lattice9}. It is noticeable that $T_c^\chi$ and $T_c^\Phi$ have different decreasing slop when magnetic field grows. The (pseudo-)critical temperature of chiral restoration $T_c^\chi$ is influenced by the magnetic field and parameter $T_0$. $T_c^\chi$ becomes larger with stronger magnetic field, but becomes smaller with smaller value of $T_0$. Their competition leads to a slow decreasing slop for $T_c^\chi$. However, $T_c^\Phi$ is not sensitive to the magnetic field but decreases as $T_0$ decreases. Therefore, we observe a faster decrease of $T_c^\Phi$.

When we obtain the reduction of (pseudo-)critical temperatures $T_c^\chi,\ T_c^\Phi$ of chiral restoration and deconfinement phase transitions, such as in the case of $T^{(2)}_0(eB)$, what is the behavior of the order parameters at finite temperature and/or magnetic field? In Fig.\ref{figmqphiimc}, the chiral condensate $\sigma/\sigma_0$ (upper panel) and Polyakov loop $\Phi$ (lower panel) are depicted as functions of temperature with fixed magnetic field $eB/m^2_\pi=0,\ 10,\ 20$ and $T^{(2)}_0(eB)$. Here, $\sigma_0$ is the chiral condensate in vacuum with vanishing temperature, density and magnetic field. With fixed magnetic field, the chiral condensate $\sigma/\sigma_0$ decreases with temperature, which indicates the restoration of chiral symmetry, and Polyakov loop $\Phi$ increases with temperature, which means the occurrence of deconfinement. At low temperature region, the chiral condensate $\sigma/\sigma_0$ increases with magnetic fields, which shows magnetic catalysis phenomena, but at high temperature region, the chiral condensate $\sigma/\sigma_0$ decreases with magnetic fields, which shows inverse magnetic catalysis phenomena. The Polyakov loop $\Phi$ increases with magnetic fields in the whole temperature region. These properties are consistent with LQCD simulations~\cite{lattice1,lattice2,lattice4,lattice5,lattice6,lattice7,lattice9}.\\

\section{three-flavor PNJL model}
\label{3fframe}

\subsection{Theoretical Framework}
The three-flavor PNJL model under external magnetic field is defined through the Lagrangian density~\cite{pnjl5,pnjl6,pnjl7,pnjl8,pnjl9,pnjl10,pnjl12},
\begin{eqnarray}
\mathcal{L}&=&\bar{\psi}\left(i\gamma^{\mu}D_{\mu}-\hat{m}_0\right)\psi+\mathcal{L}_{\text S}+\mathcal{L}_{6}-{\cal U}(\Phi,\bar\Phi),\\
\mathcal{L}_{\text S}&=&G\sum_{\alpha=0}^{8}\left[(\bar{\psi}\lambda_{\alpha}\psi)^2+(\bar{\psi}i\gamma_5\lambda_{\alpha}\psi)^2\right],\nonumber \\
\mathcal{L}_{6}&=&-K\left[\text{det}\bar{\psi}(1+\gamma_5)\psi+\text{det}\bar{\psi}(1-\gamma_5)\psi \right],\nonumber\\
{\cal U}(\Phi,{\bar \Phi}) &=&  T^4 \left[-{b_2(t)\over 2} \bar\Phi\Phi -{b_3\over 6}\left({\bar\Phi}^3+\Phi^3\right)+{b_4\over 4}\left(\bar\Phi\Phi\right)^2\right].\nonumber
\label{lagrangian}
\end{eqnarray}
The covariant derivative $D^\mu=\partial^\mu+i Q A^\mu-i {\cal A}^\mu$ couples quarks to the two external fields, the magnetic field ${\bf B}=\nabla\times{\bf A}$ and the temporal gluon field  ${\cal A}^\mu=\delta^\mu_0 {\cal A}^0$ with ${\cal A}^0=g{\cal A}^0_a \lambda_a/2=-i{\cal A}_4$ in Euclidean space. The gauge coupling $g$ is combined with the SU(3) gauge field ${\cal A}^0_a(x)$ to define ${\cal A}^\mu(x)$, and $\lambda_a$ are the Gell-Mann matrices in color space. We consider magnetic field ${\bf B}=(0, 0, B)$ along the $z$-axis by setting $A_\mu=(0,0,x B,0)$ in Landau gauge, which couples quarks of electric charge $Q=\text{diag}(Q_u,Q_d,Q_s)=\text{diag}(2/3 e,-1/3 e,-1/3 e)$. $\hat{m}_0=\text{diag}(m^u_0,m_0^d,m_0^s)$ is the current quark mass matrix in flavor space. The four-fermion interaction $\mathcal{L}_{\text S}$ represents the interaction in scalar and pseudo-scalar channels, with Gell-Mann matrices $\lambda_{\alpha},\ \alpha=1,2,...,8$ and $\lambda_0=\sqrt{2/3} \mathbf{I}$ in flavor space. The six-fermion interaction or Kobayashi-Maskawa-'t Hooft term $\mathcal{L}_{6}$ is related to the $U_A(1)$ anomaly~\cite{tHooft1,tHooft2,tHooft3,tHooft4}. The Polyakov potential ${\cal U}(\Phi,\bar\Phi)$ describes deconfinement at finite temperature, where $\Phi$ is the trace of the Polyakov loop $\Phi=\left({\text {Tr}}_c L \right)/N_c$, with $L({\bf x})={\cal P} \text {exp}[i \int^\beta_0 d \tau A_4({\bf x},\tau)]= \text {exp}[i \beta A_4 ]$ and $\beta=1/T$, the coefficient $b_2(t)=a_0+a_1 t+a_2 t^2+a_3 t^3$ with $t=T_0/T$ is temperature dependent, and the other coefficients $b_3$ and $b_4$ are constants.

It is useful to convert the six-fermion interaction into an effective four-fermion interaction in the mean field approximation, and the Lagrangian density can be rewritten as~\cite{klevansky1}
	\begin{eqnarray}
		\mathcal{L}&=&\bar{\psi}\left(i\gamma^{\mu}D_{\mu}-\hat{m}_0\right)\psi-{\cal U}(\Phi,\bar\Phi) \\
		&+&\sum_{a=0}^{8}\left[K_a^-\left(\bar{\psi}\lambda^a\psi\right)^2+K_a^+\left(\bar{\psi}i\gamma_5\lambda^a\psi\right)^2\right]\nonumber\\	&+&K_{30}^-\left(\bar{\psi}\lambda^3\psi\right)\left(\bar{\psi}\lambda^0\psi\right)+K_{30}^+\left(\bar{\psi}i\gamma_5\lambda^3\psi\right)\left(\bar{\psi}i\gamma_5\lambda^0\psi\right)\nonumber\\	&+&K_{03}^-\left(\bar{\psi}\lambda^0\psi\right)\left(\bar{\psi}\lambda^3\psi\right)+K_{03}^+\left(\bar{\psi}i\gamma_5\lambda^0\psi\right)\left(\bar{\psi}i\gamma_5\lambda^3\psi\right)\nonumber\\	&+&K_{80}^-\left(\bar{\psi}\lambda^8\psi\right)\left(\bar{\psi}\lambda^0\psi\right)+K_{80}^+\left(\bar{\psi}i\gamma_5\lambda^8\psi\right)\left(\bar{\psi}i\gamma_5\lambda^0\psi\right)\nonumber\\	&+&K_{08}^-\left(\bar{\psi}\lambda^0\psi\right)\left(\bar{\psi}\lambda^8\psi\right)+K_{08}^+\left(\bar{\psi}i\gamma_5\lambda^0\psi\right)\left(\bar{\psi}i\gamma_5\lambda^8\psi\right)\nonumber\\	&+&K_{83}^-\left(\bar{\psi}\lambda^8\psi\right)\left(\bar{\psi}\lambda^3\psi\right)+K_{83}^+\left(\bar{\psi}i\gamma_5\lambda^8\psi\right)\left(\bar{\psi}i\gamma_5\lambda^3\psi\right)\nonumber\\	&+&K_{38}^-\left(\bar{\psi}\lambda^3\psi\right)\left(\bar{\psi}\lambda^8\psi\right)+K_{38}^+\left(\bar{\psi}i\gamma_5\lambda^3\psi\right)\left(\bar{\psi}i\gamma_5\lambda^8\psi\right)\nonumber,
		\label{semilagrangian}
	\end{eqnarray}
	with the effective coupling constants
	\begin{eqnarray}
		\label{constants}
		&&K_0^\pm=G\pm\frac{1}{3}K\left(\sigma_u+\sigma_d+\sigma_s\right),\\
		&&K_1^\pm=K_2^\pm=K_3^\pm=G\mp\frac{1}{2}K\sigma_s,\nonumber\\
		&&K_4^\pm=K_5^\pm=G\mp\frac{1}{2}K\sigma_d,\nonumber\\
		&&K_6^\pm=K_7^\pm=G\mp\frac{1}{2}K\sigma_u,\nonumber\\
		&&K_8^\pm=G\mp\frac{1}{6}K\left(2\sigma_u+2\sigma_d-\sigma_s\right),\nonumber\\
		&&K_{03}^\pm=K_{30}^\pm=\pm\frac{1}{2\sqrt{6}}K\left(\sigma_u-\sigma_d\right),\nonumber\\
		&&K_{08}^\pm=K_{80}^\pm=\mp\frac{\sqrt{2}}{12}K\left(\sigma_u+\sigma_d-2\sigma_s\right),\nonumber\\
		&&K_{38}^\pm=K_{83}^\pm=\mp\frac{1}{2\sqrt{3}}K\left(\sigma_u-\sigma_d\right),	\nonumber
	\end{eqnarray}
	and chiral condensates
	\begin{eqnarray}
		\sigma_u=\langle\bar{u}u\rangle, \  \sigma_d=\langle\bar{d}d\rangle, \  \sigma_s=\langle\bar{s}s\rangle.
	\end{eqnarray}
	
The thermodynamic potential in mean field level contains the mean field part and quark part
	\begin{eqnarray}
		\label{omega1}
		\Omega_{\text {mf}} &=&2 G(\sigma_u^2+\sigma_d^2+\sigma_s^2)-4K\sigma_u\sigma_d\sigma_s+{\cal U}(\Phi,\bar\Phi)+\Omega_q,\nonumber \\
		\Omega_q &=& - \sum_{f=u,d,s}\frac{|Q_f B|}{2\pi}\sum_{l}\alpha_l \int \frac{d p_z}{2\pi} \Bigg[3E_f \nonumber\\
		&+& T\ln\left(1+3\Phi e^{-\beta E_f}+3{\bar \Phi}e^{-2\beta E_f}+e^{-3\beta E_f}\right)\nonumber\\
&+& T\ln\left(1+3{\bar \Phi} e^{-\beta E_f}+3{ \Phi}e^{-2\beta E_f}+e^{-3\beta E_f}\right)\Bigg],\nonumber
	\end{eqnarray}
	with quark energy $E_f=\sqrt{p^2_z+2 l |Q_f B|+m_f^2}$ of flavor $f=u,d,s$, longitudinal momentum $p_z$,  Landau level $l$, and effective quark masses $m_u=m_0^u-4G\sigma_u+2K\sigma_d\sigma_s$, $m_d=m_0^d-4G\sigma_d+2K\sigma_u\sigma_s$, $m_s=m_0^s-4G\sigma_s+2K\sigma_u\sigma_d$, and the degeneracy of Landau levels $\alpha_l=2-\delta_{l0}$. The ground state is determined by minimizing the thermodynamic potential,
\begin{eqnarray}
\label{gapeqs}
 \frac{\partial\Omega_{\text {mf}}}{\partial \sigma_i}=0,\ i=u,d,s,\nonumber\\
 \frac{\partial\Omega_{\text {mf}}}{\partial \Phi}=0,\ \frac{\partial\Omega_{\text {mf}}}{\partial {\bar \Phi}}=0,
\end{eqnarray}
which lead to five coupled gap equations for the order parameters $\sigma_i$, $\Phi$ and ${\bar \Phi}$. Note that there is $\Phi=\bar\Phi$ at vanishing baryon density.

Because of the contact interaction in NJL model, the ultraviolet divergence cannot be eliminated through renormalization, and a proper regularization scheme is needed. In this part, we also apply the covariant Pauli-Villars regularization~\cite{mao}. By fitting the physical quantities, pion mass $m_{\pi}=138\text{MeV}$, pion decay constant $f_{\pi}=93\text{MeV}$, kaon mass $m_K=495.7\text{MeV}$, $\eta'$ meson mass $m_{\eta\prime}=957.5\text{MeV}$ in vacuum, we fix the current masses of light quarks $m_0^{u}=m_0^{d}=5.5\text{MeV}$, and obtain the parameters $m_0^s=154.7\text{MeV}$, $G\Lambda^2=3.627$, $K\Lambda^5=92.835$, $\Lambda=1101\text{MeV}$~\cite{mei3njl}. For the Polyakov potential, its temperature dependence is fitted from the lattice simulation, and the parameters are chosen as~\cite{pnjl6} $a_0=6.75$, $a_1=-1.95$, $a_2=2.625$, $a_3=-7.44$, $b_3=0.75$, $b_4=7.5$ and $T_0=270$ MeV.

\subsection{Numerical Results}

\begin{table}
\begin{tabular}{ccccccc}
\hline
\hline
$eB$ & $T_0(eB)$ & $T_c^u$ & $T_c^d$ & $T_c^s$ & $T_c^\Phi$ \\
($m^2_{\pi}$)& ({\text {MeV}})& ({\text {MeV}})& ({\text {MeV}})& ({\text {MeV}})& ({\text {MeV}})\\
\hline
0 & 270 & 215 & 215 &268 & 216 \\
\hline
10 & 255 & 212 & 211 & 261 & 207\\
\hline
20 & 230 & 209 & 208 & 253 & 192\\
\hline
30 & 190 & 205 & 203 & 241 & 160\\
\hline
\hline
\end{tabular}
\caption{Results of (pseudo-)critical temperatures for chiral restoration and deconfinement phase transitions under external magnetic field with running parameter $T_0(eB)$.}
\label{table3pnjl}
\end{table}

\begin{figure*}[htb]
\centering
\includegraphics[width=7.5cm]{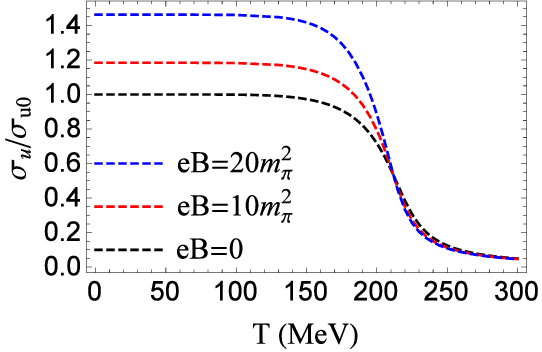}
\includegraphics[width=7.5cm]{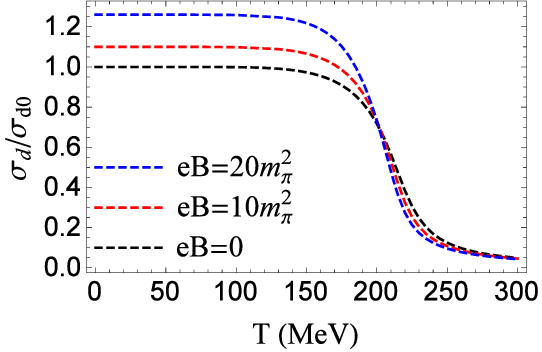}\\
\includegraphics[width=7.5cm]{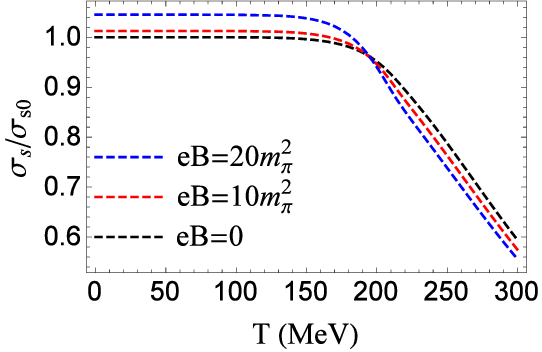}
\includegraphics[width=7.5cm]{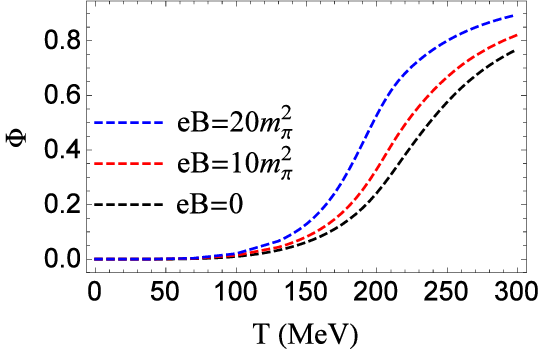}
\caption{The chiral condensates $\sigma_u/\sigma_{u0},\ \sigma_d/\sigma_{d0},\ \sigma_s/\sigma_{s0}$ and Polyakov loop $\Phi$ as functions of temperature with fixed magnetic field $eB/m^2_\pi=0,\ 10,\ 20$ and $T_0^{(2)}(eB)$. Here, $\sigma_{u0},\ \sigma_{d0},\ \sigma_{s0}$ means up, down, strange quark chiral condensate in vacuum with vanishing temperature, density and magnetic field, respectively.}
\label{figmqphi3pnjl}
\end{figure*}

As in the two-flavor PNJL model, we introduce a running Polyakov loop parameter $T_0(eB)$ in the three-flavor PNJL model, and consider its effect on the chiral restoration and deconfinement phase transitions. In the following numerical calculations, we use $T_0(eB)=T_0^{(2)}(eB)$, as shown in the red line of Fig.\ref{figtct0eb270} upper panel.

The (pseudo-)critical temperatures of chiral restoration and deconfinement phase transitions under external magnetic field with running parameter $T_0(eB)$ are listed in Tab.\ref{table3pnjl}. In nonchiral limit, the chiral restoration and deconfinement phase transitions at finite temperature and magnetic field are smooth crossover. The (pseudo-)critical temperature of chiral restoration $T_{c}^f,\ f=u,d,s$ is usually defined through the vanishing second derivative of the chiral condensate, $\frac{\partial^2 \sigma_f}{\partial T^2}$=0. The (pseudo-)critical temperature of deconfinement $T_{c}^\Phi$ is defined by the vanishing second derivative of the Polyakov loop, $\frac{\partial^2 \Phi}{\partial T^2}$=0. All the (pseudo-)critical temperatures $T_c^{u,d,s}$ and $T_c^\Phi$ decrease with magnetic fields, which is qualitatively consistent with LQCD results~\cite{lattice1,lattice2,lattice4,lattice5,lattice6,lattice7,lattice9}. Due to different electric charges of $u$ and $d$ quarks, the corresponding (pseudo-)critical temperatures $T_c^u$ and $T_c^d$ under external magnetic field are slightly different from each other ($T_c^d \leq T_c^u$). The (pseudo-)critical temperature $T_c^s$ is larger than $T_c^u$ and $T_c^d$ ($T_c^d \leq T_c^u < T_c^s$), which is caused by its heavier current quark mass. The (pseudo-)critical temperature $T_c^\Phi$ of deconfinement phase transition is close to the $T_c^u$ and $T_c^d$ in weak magnetic field cases, but becomes split with $T_c^u$ and $T_c^d$ in strong magnetic field cases. The decreasing slop of $T_c^\Phi$ is larger than  $T_c^u$, $T_c^d$ and $T_c^s$.

Figure \ref{figmqphi3pnjl} shows the chiral condensates $\sigma_u/\sigma_{u0},\ \sigma_d/\sigma_{d0},\ \sigma_s/\sigma_{s0}$ and Polyakov loop $\Phi$ as functions of temperature with fixed magnetic field $eB/m^2_\pi=0,\ 10,\ 20$ and $T_0^{(2)}(eB)$. Here, $\sigma_{u0},\ \sigma_{d0},\ \sigma_{s0}$ means up, down, strange quark chiral condensate in vacuum with vanishing temperature, density and magnetic field, respectively. With fixed magnetic field, the chiral condensates $\sigma_u/\sigma_{u0},\ \sigma_d/\sigma_{d0},\ \sigma_s/\sigma_{s0}$ decrease with temperature, which demonstrate the (partial) restoration of chiral symmetry, and the Polyakov loop $\Phi$ increases with temperature, which indicates the deconfinement process. With fixed temperature, the chiral condensates $\sigma_u/\sigma_{u0},\ \sigma_d/\sigma_{d0},\ \sigma_s/\sigma_{s0}$ increases with magnetic fields in low temperature region, which is the magnetic catalysis phenomena, but in high temperature region, they decrease with magnetic fields, which is the inverse magnetic catalysis phenomena. Polyakov loop $\Phi$ increases with magnetic fields in the whole temperature region. The results of $u$ and $d$ quark condensates and Polyakov loop $\Phi$ are consistent with LQCD results~\cite{lattice1,lattice2,lattice4,lattice5,lattice6,lattice7,lattice9}, but LQCD results of magnetic catalysis effect for $s$ quark condensate in the whole temperature region~\cite{lattice7,lattice8} can not be reproduced here.\\

\section{summary}
\label{summary}
We investigate the chiral restoration and deconfinement phase transitions under external magnetic field in terms of Pauli-Villars regularized two-flavor and three-flavor PNJL models. To mimic the reaction of the gluon sector to the presence of magnetic fields, we introduce the running Polyakov loop scale parameter $T_0(eB)$. It's found that a fast decreasing $T_0(eB)$ with the magnetic field leads to the inverse magnetic catalysis phenomena of chiral condensates of $u, d, s$ quarks, increase of Polyakov loop and the reduction of (pseudo-)critical temperatures of chiral restoration and deconfinement phase transitions.

All these results are qualitatively consistent with LQCD simulations, except for the chiral condensate of $s$ quarks. It should also be mentioned that our conclusions are qualitatively independent of the choice of logarithmic or polynomial form of Polyakov potential and the value of parameter $T_0(eB=0)$ at vanishing magnetic field. The different results of our three-flavor PNJL model with Ref\cite{pnjl3} is attributed to the different regularization scheme applied.

\noindent {\bf Acknowledgement:} The work is supported by the NSFC grant 12275204.

\end{document}